# Passive Touch Experience with Virtual Doctor Fish Using Ultrasound Haptics

Manato Yo[1], Atsushi Mastubayashi[2], Yasutoshi Makino[2] and Hiroyuki Shinoda[2]

[1] *Department of Mathematical Engineering and Information Physics, The University of Tokyo*

*7-3-1 Hongo, Bunkyo-ku, Tokyo, 113-8656 Japan*

[2] *Graduate School of Information Science and Technology, The University of Tokyo*

*7-3-1 Hongo, Bunkyo-ku, Tokyo, 113-8656 Japan*

(Email: manato-y@g.ecc.u-tokyo.ac.jp)

**Abstract ---** This study implements and evaluates passive interaction using an autostereoscopic display and ultrasound haptics technology, simulating Garra rufa ("doctor fish") nibbling. When the virtual doctor fish touches the user's hand, ultrasound tactile sensations are presented by spatio-temporal modulation (STM) as an ultrasound focal point orbits around the contact points. A user study evaluated parameters affecting realism, including STM frequency, the number of focal points, ultrasound amplitude, and hand moistening. Comparing several combinations of parameters revealed that representing contact with fewer representative points and setting the frequency of the STM to 10 Hz produced the most realistic experience.

**Keywords:** Passive Interaction, Ultrasound haptics.

## 1 INTRODUCTION

Haptic interaction with virtual objects in 3D space has great potential for entertainment applications. Ultrasound haptics technology presents tactile sensations in mid-air by focusing ultrasound waves emitted from a transducer array, which, combined with an autostereoscopic 3D display, can provide a virtual haptic interaction experience without needing to wear any device. Previous systems combining Ultrasound haptics with 3D images [1-3] have primarily focused on interaction through active touch, in which the user must actively reach out and touch the virtual object. There has been little exploration of scenarios where virtual objects approach or touch the user autonomously. Such experiences are expected to provide users with fundamentally different experiences from traditional interactions.

In this study, as a preliminary step to investigate the effects of passive interaction, we implemented a passive tactile experience system and explored the optimal parameters for tactile presentation. We implemented a system that reproduces the experience of Garra rufa ("doctor fish") as an example of passive interaction (see Fig.1). In this system, when a Garra rufa displayed on a stereoscopic display approaches and nibbles on human skin, ultrasound waves converge on that point to provide a tactile sensation. When multiple Garra rufa nibble, the ultrasound focus moves at high speed over these points, producing tactile sensations at multiple points. In this method, how multiple tactile presentation points are set, and the frequency of the focus movement affect the reality of the tactile experience. In addition, presenting a cold sensation using heat of vaporization [4] by ultrasound could improve the reality of the tactile experience. We compared these conditions in a user study to find the optimal combination of parameters.

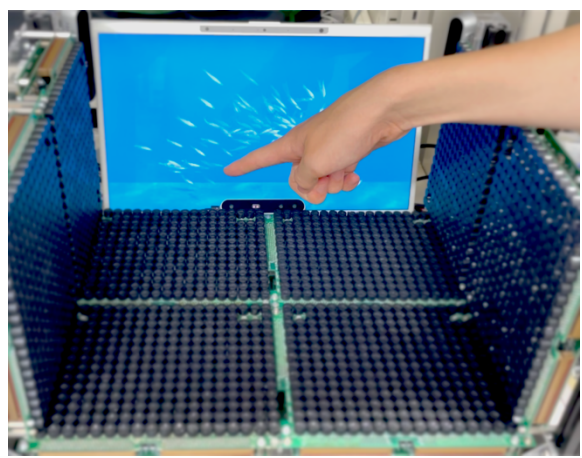

Fig.1  The visual image of the implemented system

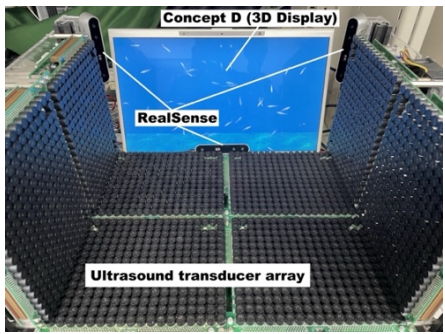

Fig.2  Setup of the system

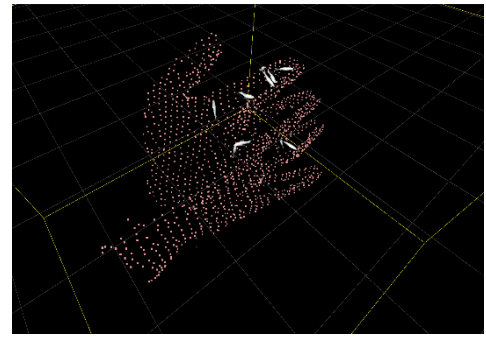

Fig.4  Point cloud of hand surface obtained by RealSense in Unity.

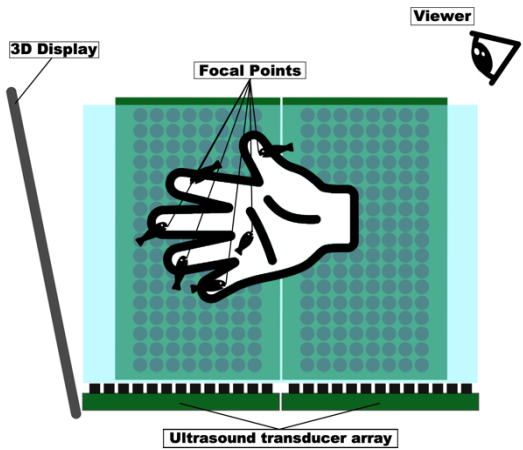

Fig.3  Illustration of how to use the system

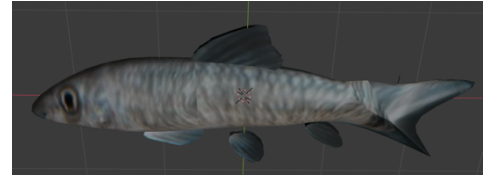

Fig.5  Screenshot of Garra rufa model in Blender.

## 2 PASSIVE INTERACTION WITH VIRTUAL GARRA RUFA

### 2.1 System Setup

We developed an experimental setup using Concept D™ for visualizing 3D images, RealSense™ for point cloud data acquisition representing the user's hand surface, and an ultrasound transducer array [5] for haptic feedback (see Fig. 2, Fig. 3). Unity™ was used for image rendering and interaction design. When a hand of a user enters the experience region (virtual aquarium), the point cloud data of the hand surface is sent to Unity and each point is converted into Unity's coordinate (see Fig. 4).

**Fish Modeling**
We created a Garra rufa model (see Fig. 5) in Blender™, incorporating two animations to reflect different swimming behaviors: one for swimming towards the hand and the other for simply patrolling. When the number of points of the hand surface exceeds the specified threshold, the fish recognizes that the hand entered the region and switches to the prowling behavior. The fish move towards the nearest point in the point cloud representing the hand surface. A fish nibbles when the distance from the nearest point is less than the specified value. Fish are programmed to swim only forward and rotate gradually, which contributes to improving realism. These calculations, such as searching the nearest point, are executed at every frame so the system is laggy when too many fish are in the field. Therefore, we use 50 fish, which ensures the system works smoothly and the number of tactile stimuli is not few.

### 2.2 Haptic Feedback

When fish nibble, ultrasound tactile sensations are presented to that point. spatio-temporal modulation (STM) [6] presented tactile sensations at multiple points. STM is a method of presenting tactile sensations at various points by periodically switching the position at which a single focus is generated. It can provide stimuli of higher intensity than methods that generate multiple focal points simultaneously.

The tactile sensation of the STM varies depending on parameters such as the position of the focal point and its switching speed. How the parameters are specified according to the biting situation of the fish has an impact on the reality of the tactile sensation. We compared the reality while varying the following parameters through a user study.

**Spatio-Temporal Modulation (STM) frequency**
The STM frequency determines the speed at which focal points are switched. This frequency significantly impacts the perceived experience. The intensity and localization of the stimulus depend on the frequency of the vibration produced on the skin. It is preferable to match the frequency of fish nibbling, but a slower STM could result in a decrease in reality due to the perceived movement of the focal point.

**Number of focal points**
Because the ultrasound focus has a radius of 5 mm to 10

mm, it does not make sense to present stimuli at several adjacent points in STM. In addition, ultrasound waves emitted by the transducer array have a reduced output during fast switching of phases [7], therefore, switching of many points should be avoided. To reduce the number of points, a clustering algorithm was employed to group nearby focal points. Focal points within a certain distance are considered a single cluster, and the focus is presented at the center of this cluster.

By reducing the number of points, the presentation time for each focus increases, leading to a stronger perceived stimulus. However, excessively reducing the number of points may result in a perceived lack of stimulation relative to the visual image.

**Ultrasound amplitude**

The amplitude of the output ultrasound wave of the transducer array is directly related to the intensity of the stimulus. Given the small size of Garra rufa, an excessively strong stimulus would detract from the realism of the experience. Therefore, it is crucial to select an appropriate amplitude.

**Hand moistening**

Since Garra rufa lives in water, we hypothesized that wetting the hand would enhance the realism of the experience. The presentation of a cooling sensation due to the vaporization of water by ultrasound focusing has the potential to enhance the reality of the experience. In the following experiment, a spray bottle is used to moisten the hand with room-temperature tap water. To prevent water damage to the devices, we lightly mist the hand, ensuring only minimal moisture is applied.

## 3 USER STUDY

### 3.1 Procedure

We conducted a user study with 6 male participants, only one of whom has experience with real doctor fish. We evaluated 5 different parameter combinations. Given the impracticality of testing all parameter combinations, we first established a baseline (Condition 1). We then prepared the other four conditions by varying parameters from this baseline. Each new condition was designed to be distinctly different from the baseline but with the potential to improve the experience. Condition 1 serves as the baseline, with an STM frequency of 2 Hz, no clustering method applied, and no moistening. Ultrasound transducers are driven at the maximum amplitude. In Condition 2, only the STM frequency is increased to 10 Hz while the other parameters remain the same as the baseline. Condition 3 maintains the increased STM frequency of 10 Hz and introduces a clustering method, which results in an average of approximately 20 haptic points. In Condition 4, only the amplitude of the ultrasound is reduced by a factor of 0.75. In Condition 5, the hand is moistened while all other parameters remain at baseline values. The conditions of this experiment were presented in a fixed order for all participants.

The participants rated the realism of the experience on a 5-point scale after 1-2 minutes of experience on each condition. Additionally, they evaluated the perceived match between visual and tactile stimuli on a 5-point scale, where 1 indicated too few stimuli, 3 indicated a good match, and 5 indicated too many stimuli. They are asked to evaluate both the experience with the whole hand and the one with only the index finger. This is to investigate the difference in experience depending on the density of tactile stimuli. Finally, as evaluations regarding the images, participants responded on a 4-point scale to the questions "The fish seemed realistic," "The fish seemed to pop out of the screen," and "The position of the tactile stimuli matched with the images, where "not at all" was 1, and "completely" was 4. The purpose of this was to confirm whether the 3D images were appropriately viewed by the participants.

### 3.2 Results

Table 1 The result of the experiment (tactile)

| Condition | Reality (1~5) | Match between the number of visual and tactile stimuli (1~5) | |
|---|---|---|---|
| | | Whole hand | Index finger |
| 1 | 4.33 | 2.83 | 3.33 |
| 2 | 3.67 | 3.33 | 3.50 |
| 3 | 4.50 | 3.17 | 3.00 |
| 4 | 2.00 | 1.83 | 1.50 |
| 5 | 4.17 | 2.83 | 2.67 |

Table 2 The result of the experiment (visual)

| Visual reality (1~4) | Fish pop up (1~4) | Match between the location of visual and tactile stimuli (1~4) |
|---|---|---|
| 3.50 | 3.33 | 3.50 |

Table 1 shows the result of the experiment. Each value is an average value of participants' response. Condition 3, in which STM frequency is increased but the focal points are reduced, got the highest score. This result revealed that the perceived number of tactile stimuli does not always correspond to the actual number of focal points. The condition in which reduced the number of focal points while increasing STM frequency, was rated as the most matched experience between visual and tactile stimuli. In addition, small ultrasound amplitude reduces the number of perceived focal points.

The effect of hand moistening on the perceived realism was inconclusive, with mixed responses from participants. A participant reported that moistening the hand significantly enhanced the sense of realism. However, another participant had the opposite experience, stating that moistening the hand felt artificial thus detracted from the overall realism.

Table 2 shows that the visuality of the introduced system was highly appreciated and that the subjects were able to experience this system appropriately. Each value on this table is an average value, noting that the max value is 4.

### 3.3 Limitations and Future Work

The sample size was relatively small, and all participants were male. Future studies should include a larger, more diverse participant pool.

Future work should explore:
- Comparison between active and passive interactions especially on how the attention towards users affects the experiences.
- Introduction of constant airflow to enhance the underwater sensation.
- Application of this technology to other scenarios beyond the Garra rufa simulation.

## 4 Conclusion

This study demonstrates the potential of combining ultrasound haptics with 3D displays for creating realistic passive interactions in virtual environments. By carefully tuning haptic parameters and ensuring tight integration with visual feedback, we can create compelling virtual experiences that engage multiple senses. Our findings have implications for the development of virtual and augmented reality systems, particularly in creating more immersive and realistic haptic experiences. The successful implementation of passive interaction opens new possibilities in fields such as entertainment, education, and therapeutic applications.